\documentclass[11pt]{article}
\usepackage{amsmath,amssymb,amsthm}
\usepackage{hyperref}
\usepackage{amsthm}
\usepackage{graphicx}
\usepackage{color}
\usepackage[normalem]{ulem}
\usepackage{authblk}

\graphicspath{{../Figures/}}

\theoremstyle{plain}

\theoremstyle{definition}

\definecolor{lightblue}{rgb}{0.30,0.30,1}	

\begin{document}

\title{Is there sufficient evidence for criticality in cortical systems?}

\author[1]{Alain Destexhe}
\author[2]{Jonathan D. Touboul}
\affil[1]{Paris-Saclay Institute of Neuroscience, CNRS, Gif sur Yvette, France}
\affil[2]{Department of Mathematics and Volen Center for Complex Systems, Brandeis University.}

\date{\today}%

\maketitle
\begin{center}
	\begin{minipage}{0.75\textwidth}
\textbf{Abstract} \emph{Many studies have found evidence that the brain operates at a critical
point, a processus known as self-organized criticality.  A recent paper 
found remarkable scalings suggestive of criticality in systems as different
as neural cultures, anesthetized or awake brains.  We point out here that
the diversity of these states would question any claimed role of criticality 
in information processing.  Furthermore, we show that two non-critical 
systems pass all the tests for criticality, a control that was not provided
in the original article.  We conclude that such false positives demonstrate 
that the presence of criticality in the brain is still not proven and that we
need better methods that scaling analyses.}
\end{minipage}
\end{center}

Whether the brain operates at criticality or not has been debated in the past almost 20 years since the original report by Beggs and Plenz~\cite{beggs} of power-law statistics of some aspects of firing patterns that can evoke the behavior of specific physical systems at a transition between two states. This suggestion sparked the interest of theoreticians, in search for a rigorous validation of this hypothesis and for a theory of the origin and implications of criticality in neural systems~\cite{mora2011biological}, and experimentalists in search for evidence of criticality in various neural systems and various physiological or pathological brain states~\cite{friedman2012universal,hahn2010neuronal,massobrio2015criticality}; some papers even hypothesized that criticality was hallmark of healthy brain function and optimal information capacity~\cite{shew2013functional,shew2009neuronal,beggs2007criticality}. 

However, many theoretical works showed that the evidence provided for criticality in experiments are not specific to critical systems, and that simple phenomena, assuming much less structure in the brain, could play a role in the emergence of the hallmarks of criticality used, including artifacts of thresholding noisy signals~\cite{touboul-destexhe:10}, or the result of a intermittent activity, as observed in the middle of past century~\cite{miller1957some} and that may affect high-dimensional neural data~\cite{aitchison2014zipf} or large-scale interacting networks~\cite{touboul,schwab2014zipf}. One of the key difficulty to date related to the criticality hypothesis is the lack of a univocal statistical test, which, to some extent, lends itself to the difficulty to point to a specific type of phase transition associated with the putative critical dynamics. 

With the aim of bringing together theoretical findings and experimental data, we investigate the conclusions of a recent paper reporting remarkable power-law scalings on an extensive dataset, allowing to analyze with identical methods with species, distinct preparations and different brain states~\cite{fontenele}. Quite strikingly, data from freely moving or anesthetized mammals to ex-vivo preparations of reptile nervous system or cultured slices of rat cortex all show common scaling in a specific activity regime that the authors interpret as a unspecified critical regime (said to be distinct from the classical mean-field directed percolation model). This raises at least two theoretical questions: (i) are evidence sufficient to claim that the regime associated is associated with a critical regime from a realistic model, and (ii) what can the extraordinary consistency of the scalings observed can indicate about such vastly distinct neural states associated with so distinct information processing properties. To assess criticality, the authors propose a test based on the relationship between power-law scaling exponents of neuronal avalanches in experiments and in a model at criticality inspired from classical crackling-noise systems. However, the authors in~\cite{fontenele} did not provide any control (non-critical systems) to assess whether the methodology distinguishes critical and non-critical models of neural networks, and in fact all the systems they considered passed the test. 

\begin{figure}
	\centerline{\includegraphics[width=.6\textwidth]{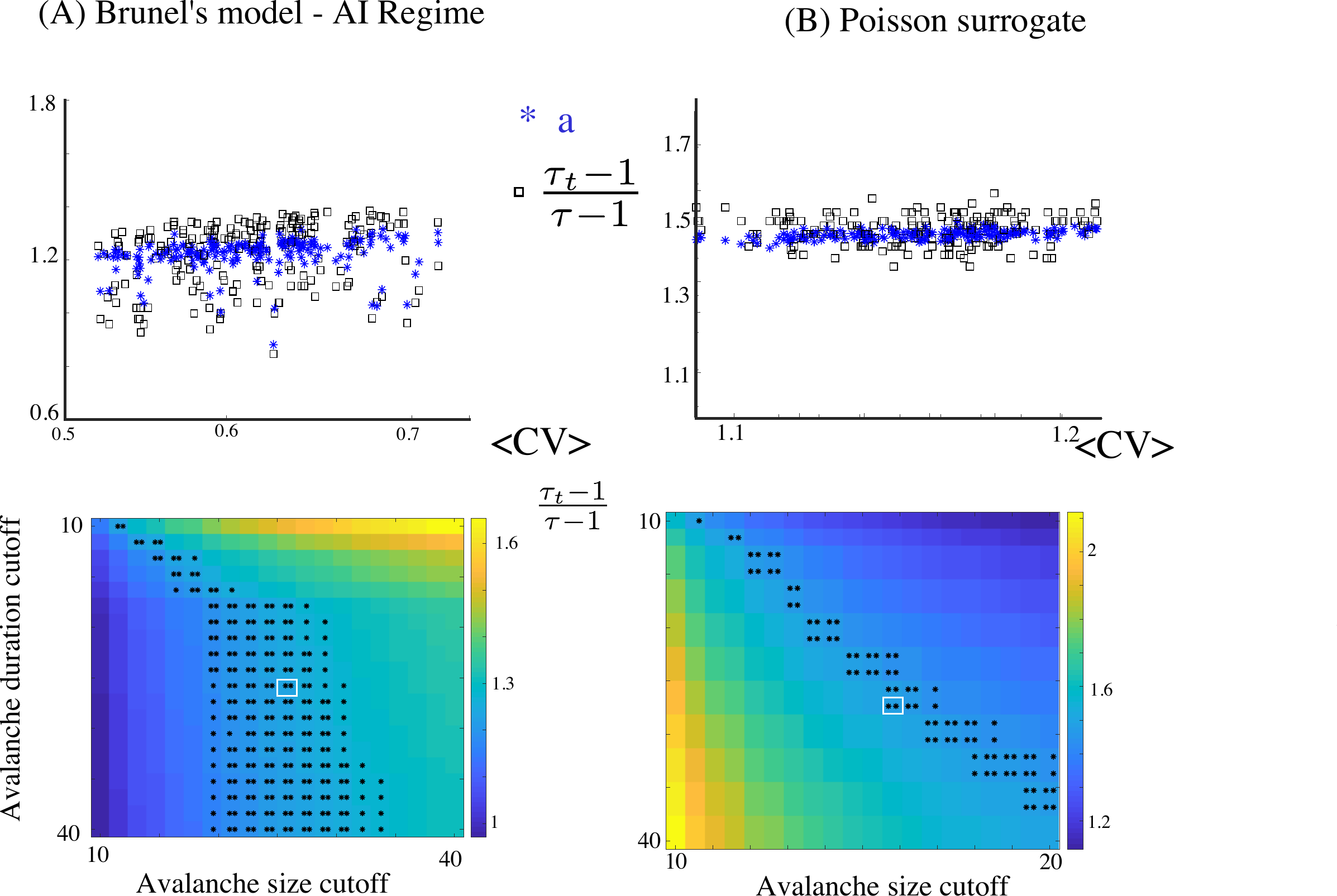}}
	\caption{200 simulations of the Brunel model (A, parameters as in~\cite[Fig 7]{touboul}) and Poisson surrogate (B, Ornstein-Uhlenbeck rate with randomly chosen coefficients). Bottom: A multitude of combinations of cutoffs yield results compatible with Sethna's relationship  (2-sample t-test comparing the distribution of ratios and $a$, *: $p<0.01$, **: $p<0.05$, for $n=14$ instances as in~\cite{fontenele}). Top: two counter examples in each case (white squares on bottom figures). }
	\label{figure}
\end{figure}

In detail, the methodology used by the authors consists in fitting with power-laws the distribution of neuronal avalanche size (exponent $\tau$) and duration (exponent $\tau_t$) truncated to a cutoff, validating the fit by comparing the Akaike information criterion (AIC) associated with the AIC of a log-normal fit. The authors found acceptable support for power-law distributions, but noted that exponents found are not compatible with the mean-field directed percolation systems generally used as a reference to assess criticality. Notwithstanding, they proposed to classify systems as critical when exponents satisfy \emph{Sethna's crackling relationship}:
\begin{equation}\label{eq:sethna}
	\frac{\tau_t-1}{\tau-1}=a
\end{equation}
where $a$ the scaling of the average avalanche size as a function of duration. Renormalization theory indeed showed that this scaling is universal, at  criticality, for a specific class of systems called \emph{crackling noise} systems. The choice of this test implicitly assumes that the neuronal systems belong to the universality class of crackling systems, which to date remains an open question and has not been established. In fact, the authors refer to our own theoretical paper~\cite{fontenele} as a support for the use of this relationship to distinguish critical from non-critical system. However, in that paper, the results were not construed to support the test performed here. In detail, we showed in that paper that for two non-critical models, all hallmarks used for criticality in the experimental literature were satisfied by classical neural network models and a surrogate network, yet, these counter-examples would not satisfy Sethna's relationship~\eqref{eq:sethna} in the thermodynamic limit and for the scaling of the \emph{tails} of the distributions of avalanches. There is no claim about any scaling related to the bulk of avalanche distributions, and moreover our results were not construed as a demonstration that systems that do satisfy~\eqref{eq:sethna} are critical, or that those for which~\eqref{eq:sethna} is not satisfied are not critical. We therefore thoroughly replicated the analysis in Fontenele in the two models studied in~\cite{touboul} (the Brunel network~\cite{brunel:00} and a stochastic surrogate~\cite{touboul}) and investigated fitting the bulk up to a cutoff affects the tests. We therefore performed extensive simulations, computed avalanche distributions for size and duration, fitted power-law distributions with various cutoffs, and used the AIC difference test proposed in~\cite{fontenele} to validate the power-law fits ($100\%$ of the $n=32\,000$ distributions considered in Fig.~\ref{figure} passed the test). We next checked, based on the fitted exponents, whether~\eqref{eq:sethna} was statistically valid using a two-sample t-test. We found that a multitude of those counter-examples are consistent with~\eqref{eq:sethna}, and would therefore be classified as critical by the criterion used in~\cite{fontenele}.

These many counter-exemples to the test used in Fontenele highlights the fact that the evidence provided is not sufficient to establish that the data they analyzed reflects criticality. Truncating the data (here, thresholds as low as 15-25 duration bins) is a good practice and inevitable for experimental datasets, yet this can affect substantially the statistics of the tails and when fits are done \emph{the smallest observable avalanche}. Therefore, while the authors do report evidence that the brain, in some regimes, shows statistics that are consistent with a given type of critical systems, they did not establish that the brain operates at criticality, because the methodology used seems insufficient to distinguish critical from non-critical systems. 

Brunel's model is a well-known simple network model displaying activity states relevant to cortical activity. The fact that it can reproduce all aspects of the analyses in~\cite{fontenele} away from criticality suggests that the most parsimonious explanation for the data would not require resorting to criticality, a regime that either entail fine tuning of physiology parameters or homeostatic mechanisms for constraining self-organization to criticality. Moreover, such basic phenomena could appear consistent with the ubiquity of these observations in a variety of neural systems from awake animals to reptile ex-vivo neurons. This report underlines once more that the criticality hypothesis is yet to be established, and that rigorous methods should be developed, experimentalists and theoreticians hand in hand, to make precise definitions of the type of criticality that could arise in brain and to establish rigorous, univocal tests for that criticality.

\bibliographystyle{plain}
\bibliography{Response}
\end{document}